\documentclass[10pt,superscriptaddress,nofootinbib,notitlepage,tightenlines,twocolumn, pra]{revtex4-1}

\usepackage{amsthm,amsmath,amssymb}

\usepackage{mathrsfs}
\usepackage{amsthm}
\usepackage{amsmath}
\usepackage{amssymb}
\usepackage{graphicx}
\usepackage{epstopdf}
\usepackage{url}
\usepackage{float}
\usepackage{bm}
\usepackage{ifthen}
\usepackage[usenames,dvipsnames]{color}
\usepackage{mathrsfs}
\usepackage[colorlinks=true,citecolor=blue,urlcolor=black]{hyperref}
\usepackage{float}

\newcommand{\be}{\begin{equation}}
\newcommand{\ee}{\end{equation}}

\newcommand{\rv}[1]{{\bf{#1}}}

\newcommand{\sket}[1]{{\ensuremath{\lvert#1\rangle}}}
\newcommand{\lket}[1]{{\ensuremath{\left\lvert#1\right\rangle}}}
\newcommand{\ket}[1]{\if@display\lket{#1}\else\sket{#1}\fi}

\newcommand{\sbra}[1]{{\ensuremath{\langle#1\rvert}}}
\newcommand{\lbra}[1]{{\ensuremath{\left\langle#1\right\rvert}}}
\newcommand{\bra}[1]{\if@display\lbra{#1}\else\sbra{#1}\fi}

\newcommand{\sbraket}[2]{{\ensuremath{\langle#1\rvert#2\rangle}}}
\newcommand{\lbraket}[2]{{\ensuremath{\left\langle#1\!\left\rvert\vphantom{#1}#2\right.\!\right\rangle}}}
\newcommand{\braket}[2]{\if@display\lbraket{#1}{#2}\else\sbraket{#1}{#2}\fi}

\newcommand{\sketbra}[2]{{\ensuremath{\lvert #1\rangle\!\langle #2\rvert}}}
\newcommand{\lketbra}[2]{{\ensuremath{\left\lvert #1\right\rangle\!\!\left\langle #2\right\rvert}}}
\newcommand{\ketbra}[2]{\if@display\lketbra{#1}{#2}\else\sketbra{#1}{#2}\fi}

\newcommand{\cX}{\mathcal{X}}

\newcommand{\cZ}{\mathcal{Z}}

\newcommand{\rvS}{\textbf{S}}

\newcommand{\sX}{\mathsf{X}}
\newcommand{\sZ}{\mathsf{Z}}

\theoremstyle{plain}

\theoremstyle{definition}

\begin{document}
\title{Tight security bounds for decoy-state quantum key distribution}
\author{Hua-Lei Yin}
\email{hlyin@nju.edu.cn}
\affiliation{National Laboratory of Solid State Microstructures and School of Physics, Nanjing University, Nanjing 210093, China}
\affiliation{Zhongchuangwei Quantum Co., Ltd., Beijing 101400, China}
\author{Min-Gang Zhou}
\author{Jie Gu}
\author{Yuan-Mei Xie}
\author{Yu-Shuo Lu}
\author{Zeng-Bing Chen}
\email{zbchen@nju.edu.cn}
\affiliation{National Laboratory of Solid State Microstructures and School of Physics, Nanjing University, Nanjing 210093, China}

\begin{abstract}
The BB84 quantum key distribution (QKD) combined with decoy-state method is currently the most practical protocol, which has been proved secure against general attacks in the finite-key regime. Thereinto, statistical fluctuation analysis methods are very important in dealing with finite-key effects, which directly affect secret key rate, secure transmission distance and even the most important security. There are two tasks of statistical fluctuation in decoy-state BB84 QKD. One is the deviation between expected value and observed value for a given expected value or observed value. The other is the deviation between phase error rate of computational basis and bit error rate of dual basis. Here, we provide the rigorous and optimal analytic formula to solve the above tasks, resulting higher secret key rate and longer secure transmission distance. Our results can be widely applied to deal with statistical fluctuation in quantum cryptography protocols.
\end{abstract}
\maketitle

\section{Introduction}
So far, there have existed many kinds of protocols describing how quantum key distribution (QKD) work, such as the Bennett-Brassard-1984 (BB84)~\cite{bennett1984proceedings}, Bennett-Brassard-Mermin-1992 ~\cite{bennett1992quantum}, Bennett-1992~\cite{bennett1992quantum1}, and six-state~\cite{bruss1998optimal} protocols.~Although different protocols contains different processes, they all serve the same purpose to guarantee that two parties, named Alice and Bob, can share a string of key data through a channel fully controlled by an eavesdropper, named Eve~\cite{bennett1984proceedings}. Unlike some computational assumptions, these protocols are all proven to be secure with fundamental physical laws in the recent years~\cite{mayers2001unconditional,lo1999unconditional,shor2000simple,tamaki2003unconditionally,renner2008security,koashi2009simple,tomamichel2011uncertainty}, which shows the great advantage in information transmitting that QKD holds.
BB84 stands out as the most important protocol due to its best overall performance.
However, implementations of the BB84 protocol differ from the original theoretical proposal. For an ideal single-photon source is not available yet, in actuality, a weak pulsed laser source is in place of it. Nevertheless, there is a critical flaw in the weak pulsed laser source that an non-negligible part of laser pulses contains more than one photon, which will be exploited by Eve through the photon-number-splitting (PNS) attack~\cite{brassard2000limitations}. To address this drawback with high channel loss, the decoy-state method is introduced~\cite{hwang2003quantum,wang2005beating,lo2005decoy}.

The source will generate the phase-randomized coherent state in decoy-state method, which can be regarded as the mixed photon number state. The essence of the decoy state idea can be summarized as that the yield (bit error rate) of $n$-photon in signal state is equal to that in decoy state.
However, this equal-yield condition can only be established under the asymptotic-key regime. The expected value of yield (bit error rate) of $n$-photon in signal state and decoy state are identical while the corresponding observed value cannot be assumed to be the same in the finite-key regime. By exploiting the decoy-state method, one can establish the linear system of equations about the expected values to obtain expected value of yield (bit error rate) of the single-photon component, where we need estimate the expected value of some parameters given by the known observed values. Actually, the observed value of yield (bit error rate) of the single-photon component in the key extraction data is what we really need, where we must estimate the observed value given by the known expected value.

The Gaussian analysis method~\cite{ma2005practical} is first proposed to deal with the the deviation between expected value and observed value given by the known observed value. The Gaussian analysis method is not rigorous because of the identically distributed assumption, which can only valid in the collective attack. Resulting the extracted secret key cannot be secure against the coherent attack.
Recently, the multiplicative form Chernoff bound~\cite{curty2014finite} and Hoeffding inequality~\cite{lim2014concise} methods are proposed to remove the identically distributed assumption, respectively. However, there is a considerable gap between the secret key rate bounds obtained from Chernoff-Hoeffding method and that obtained from the Gaussian analysis.
In order to close this gap, the inverse solution Chernoff bound method~\cite{zhang2017improved} is presented, which achieves a similar performance with Gaussian analysis. Here, we should point out that the inverse solution Chernoff bound method is also seem to be not rigorous. An important assumption in Chernoff bound is that one should have the prior knowledge of expected value. However, the problem that we have in hand is the opposite that we need to estimate expected value for a given observed value. This is why the multiplicative form Chernoff bound is somehow complex and carefully tailored. A direct criterion is that the lower bound result of inverse solution Chernoff bound is superior to the Gaussian analysis when one has a small observed value. Note that the result of Gaussian analysis should be optimal because the identically distributed assumption is a special case.

For BB84 protocol, one need bound the the conditional smooth min-entropy~\cite{tomamichel2012tight}, which relates to the phase error rate. The phase error rate cannot be directly observed, which can only be estimated by using the random sampling without replacement theory for security against the general attacks. A hypergeometric distribution method~\cite{fung2010practical} is first proposed to deal with the deviation between phase error rate of computational basis and bit error rate of dual basis in the finite-key regime. By using the inequality scaling technique, a numerical equation solution by using Shannon entropy function~\cite{fung2010practical} is acquired to estimate the phase error rate. Based on this, an analytical solution is obtained when the data size is large~\cite{lim2014concise}. A looser analytical solution is using the Serfling inequality~\cite{curty2014finite}. By exploiting the Ahrens map for Hypergeometric distribution, one uses Clopper-Pearson confidence interval~\cite{lucamarini2015security} replace the Serfling inequality.
Recently, a specifically tailored analytical solution is acquired~\cite{korzh2015provably}, which achieves a big advantage compared to Serfling inequality.
Here, we should point out that the specifically tailored analytical solution~\cite{korzh2015provably} for random sampling without replacement is incorrect. The inequality scaling of binomial coefficient and Eq. (11) in supplementary information of Ref.~\cite{korzh2015provably} is wrong.

In order to further improve the secret key rate in the case of high-loss, some authors of us have developed the tightest method to solve the above two tasks of statistical fluctuation~\cite{yin2019finite}. Thereinto, the numerical equation of Chernoff bound is used to estimate the observed value for a given expected value. A numerical equation of Chernoff bound's variant is exploited to obtain the expected value for a given observed value. A numerical equation relating to the hypergeometric distribution is directly applied to acquire the phase error rate for a given bit error rate. These numerical equation solutions are very tight but they are very inconvenient to use. On the one hand, it will be very time consuming if we optimize the system parameters globally by solving transcendental equations. On the other hand, it is a challenge to solve transcendental equations for each time post-processing in commercial QKD system with hardware. In this work we present the optimal analytical formulas to solve the two tasks of statistical fluctuation by using the rigorous inequality scaling technique. Furthermore, we establish the complete finite-key analysis for decoy-state BB84 QKD with composable security. The simulation results show that the secret key rate and secure transmission distance of our method have a significant advantage compared with previous rigorous methods.

\section{Statistical Fluctuation Analysis}
In this section, we let $x^*$ be the expected value, $x$ be the observed value, $\underline{x}$ and $\overline{x}$ be the lower and upper bound of $x$. Here, we first introduce the numerical equation result of Ref.~\cite{yin2019finite}. Then we present the tight analytical formulas by using the rigorous inequality scaling technique, which are the slightly looser bounds than those obtained by solving equations.

\subsection{Random Sampling Without Replacement}

Random sampling without replacement.---Let $X_{n+k}$$:=\{x_1,x_2,...,x_{n+k}\}$ be a string of binary bits with $n+k$ size, in which the number of bits value is unknown. Let $X_k$ be a random sample (without replacement) bit string with $k$ size from $X_{n+k}$. Let $\lambda$ be the probability of bit value 1 observed in $X_k$. Let $X_n$ be the remaining bit string, where the probability of bit value 1 observed in $X_n$ is $\chi$. Then, in this article, we let $C^j_i=\frac{i!}{j!(i-j)!}$ be the binomial coefficient. For any $\epsilon > 0$, we have the upper tail
${\rm Pr}[\chi\ge \lambda + \gamma^{U}] \le \epsilon $,
where we use $\gamma^{U}$ represents $\gamma^{U}(n,k,\lambda,\epsilon)$  and $\gamma^{U}$  is the positive root of the following equation~\cite{yin2019finite}
\begin{equation}
\begin{aligned}\label{1eq1}
\ln C_{k}^{k\lambda}+\ln C_{n}^{n(\lambda+\gamma^{U})}-\ln C_{n+k}^{(n+k)\lambda+n\gamma^{U}}=\ln \epsilon.
\end{aligned}
\end{equation}

\begin{figure}[t!]
  \centering
  \includegraphics[width=80mm]{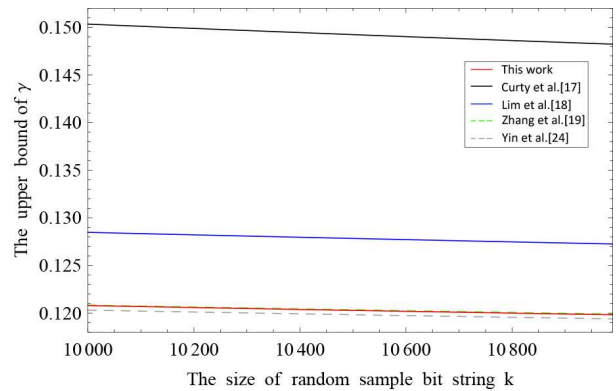}
  \caption{\textbf{}\label{f1} Comparison of the random sampling without replacement for five methods: our analytic result, analytic result with Serfling inequality~\cite{curty2014finite}, approximate analytic result~\cite{lim2014concise}, numerical result with Shannon entropy function~\cite{zhang2017improved} and optimal numerical result with binomial coefficient~\cite{yin2019finite}. Let $n=10^{5}$ and failure probablity $\epsilon=10^{-10}$.}
\end{figure}

Calculating Eq.~\eqref{1eq1}, we get numerical results of $\gamma^{U}$, corresponding to the upper bound of the random sampling without replacement.
Solving transcendental equation Eq.~\eqref{1eq1} is usually very complicated.
Here, we are going to make use of some techniques mathematically to get rigorous tight analytical result. Details proof can be found in Appendix A. For the upper tail, let $0<\lambda<\chi\leq0.5$, we have the analytical result
\begin{equation}\label{eq5}
\gamma^{U}=\frac{\frac{(1-2\lambda)AG}{n+k}+
\sqrt{\frac{A^2G^2}{(n+k)^2}+4\lambda(1-\lambda)G}}{2+2\frac{A^2G}{(n+k)^2}},
\end{equation}
where $A=\max\{n,k\}$ and $G=\frac{n+k}{nk}\ln{\frac{n+k}{2\pi nk\lambda(1-\lambda)\epsilon^{2}}}$.
Therefore, the upper bound of $\chi$ can be given by $\chi=\lambda+\gamma^{U}$ with a failure probability $\epsilon$. Fig.~\ref{f1} shows the comparison results between our method and previous method~\cite{curty2014finite,lim2014concise,zhang2017improved,yin2019finite}, which means that our analytic result is optimal and closes to the numerical results.

\subsection{Deviation Between Expected and Observed Values}

\subsubsection{For a given expected value}
Chernoff bound.---Let $X_1, X_2...,X_N$ be a set of independent Bernoulli random variables that satisfy ${\rm Pr}(X_i=1)=p_i$ (not necessarily equal), and let $X:=\sum_{i=1}^NX_i$.
The expected value of $X$ is denoted as $x^*:=E[X]=\sum_{i=1}^Np_i$. An observed value of $X$ is represented as $x$ for a given trial. Note that, we have $x\geq0$, $x^*\geq0$, $x^*$ is known and $x$ is unknown.
For any $\epsilon>0$, we have the upper tail ${\rm Pr}[x\geq(1+\delta^{U})x^*]\leq\epsilon$,
where we use $\delta^{U}$ represents $\delta^{U}(x^{*},\epsilon)$ and $\delta^{U}>0$ is the positive root of the following equation~\cite{yin2019finite}
\begin{equation}
\begin{aligned}\label{eq12}
x^*[\delta^{U}-(1+\delta^{U})\ln(1+\delta^{U})]=\ln\epsilon.
\end{aligned}
\end{equation}
For any $\epsilon>0$, we have the lower tail ${\rm Pr}[x\leq(1-\delta^{L})x^{*}]\leq\epsilon$, where we use $\delta^{L}$ represents $\delta^{L}(x^{*},\epsilon)$ and $0<\delta^{L}\leq1$ is the positive root of the following equation~\cite{yin2019finite}
\begin{equation}
\begin{aligned}\label{eq14}
-x^{*}[\delta^{L}+(1-\delta^{L})\ln(1-\delta^{L})]=\ln\epsilon.
\end{aligned}
\end{equation}

By solving  Eq.~\eqref{eq12} and Eq.~\eqref{eq14}, we get numerical results of $\delta^{U}$ and $\delta^{L}$, corresponding to the upper bound and lower bound.
Solving transcendental equations Eq.~\eqref{eq12} and Eq.~\eqref{eq14} are usually very complicated.
For the upper tail, by using the inequality $\ln(1+\delta^{U})>2\delta^{U}/(2+\delta^{U})$ in Eq.~\eqref{eq12}, we have the analytical result
\begin{equation}\label{eq17}
\delta^U=\frac{\beta+\sqrt{8\beta x^{*}+\beta^2}}{2x^{*}},
\end{equation}
where we let $\beta=\ln\epsilon^{-1}$. For the lower tail, by using the inequality $-\ln(1-\delta^{L})<\delta^{L}(2-\delta^{L})/[2(1-\delta^{L})]$ in Eq.~\eqref{eq14}, we have the analytical result
\begin{equation}\label{eq18}
\delta^L=\sqrt{\frac{2\beta}{x^{*}}}.
\end{equation}

Therefore, the lower and upper bound of observed value $x$ for a given expected value $x^*$ can be given by $\overline{x}=x^{*}+\frac{\beta}{2}+\sqrt{2\beta x^{*}+\frac{\beta^2}{4}}$ and $\underline{x}=x^{*}-\sqrt{2\beta x^{*}}$ with a failure probability $\epsilon$, respectively. Note that we must have the lower bound $\underline{x}\geq0$. The analytic result of upper bound in Eq.~\eqref{eq17} is also acquired in Ref.~\cite{zhang2017improved} while we obtain more optimal lower bound in Eq.~\eqref{eq18}.

\begin{figure}[t!]
  \centering
  \includegraphics[width=80mm]{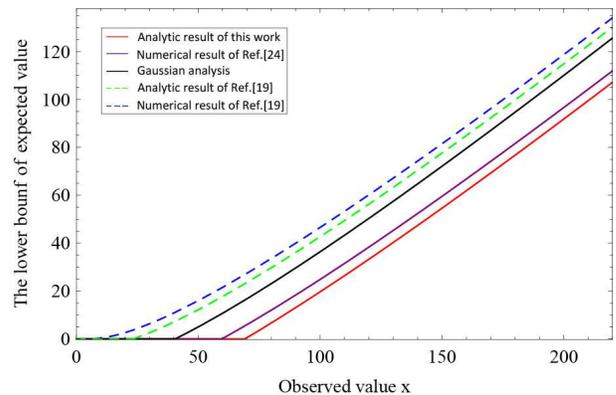}
  \caption{\textbf{}\label{f2} Comparison of the lower bound of expected value given a observed value for five methods: our analytic result, numerical result of Ref.~\cite{yin2019finite}, Gaussian analysis, numerical result and analytic result of Ref.~\cite{zhang2017improved}.}
\end{figure}

\subsubsection{For a given observed value}
Variant of Chernoff bound.---Let $X_1, X_2...,X_N$ be a set of independent Bernoulli random variables that satisfy ${\rm Pr}(X_i=1)=p_i$ (not necessarily equal), and let $X:=\sum_{i=1}^NX_i$.
The expected value of $X$ is denoted as $x^*:=E[X]=\sum_{i=1}^Np_i$. An observed outcome of $X$ is represented as $x$ for a given trial. Note that, we have $x\geq0$, $x^*\geq0$, $x$ is known and $x^*$ is unknown.
For any $\epsilon>0$, we have the upper tail ${\rm Pr}[x^{*}\leq x+\Delta^{U}]$, where we use $\Delta^{U}$ represents $\Delta^{U}(x,\epsilon)$ and $\Delta^{U}$ is the positive root of the following equation~\cite{yin2019finite}
\begin{equation}
\begin{aligned}\label{eq20}
-\Delta^{U}+x\ln\frac{x+\Delta^{U}}{x}=\ln\epsilon.
\end{aligned}
\end{equation}
For any $\epsilon>0$, we have the upper tail ${\rm Pr}[x^{*}\geq x+\Delta^{L}]$, where we use $\Delta^{L}$ represents $\Delta^{L}(x,\epsilon)$ and $\Delta^{L}$ is the positive root of the following equation~\cite{yin2019finite}
\begin{equation}
\begin{aligned}\label{eq22}
\Delta^{L}-(x+\Delta^{L})\ln\frac{x+\Delta^{L}}{x}=\ln\epsilon.
\end{aligned}
\end{equation}

By solving  Eq.~\eqref{eq20} and Eq.~\eqref{eq22}, we get numerical results of $\Delta^{U}$ and $\Delta^{L}$, corresponding to the upper bound and lower bound.
Solving transcendental equations Eq.~\eqref{eq20} and Eq.~\eqref{eq22} are usually very complicated.
For the upper tail, by using the inequality $\ln\left(1+\frac{\Delta^{U}}{x}\right)<\frac{\Delta^{U}}{x}\left(2+\frac{\Delta^{U}}{x}\right)/\left[2\left(1+\frac{\Delta^{U}}{x}\right)\right]$ in Eq.~\eqref{eq20}, we have the analytical result
\begin{equation}
\begin{aligned}\label{eq25}
\Delta^U=\beta+\sqrt{2\beta x+\beta^2}.
\end{aligned}
\end{equation}
For the lower tail, by using the inequality $\ln\left(1+\frac{\Delta^{L}}{x}\right)>2\frac{\Delta^{L}}{x}/\left(2+\frac{\Delta^{L}}{x}\right)$ in Eq.~\eqref{eq22}, we have the analytical result
\begin{equation}
\begin{aligned}\label{eq26}
\Delta^L=\frac{\beta}{2}+\sqrt{2\beta x+\frac{\beta^2}{4}}.\\
\end{aligned}
\end{equation}

Therefore, the lower and upper bound of expected value $x^{*}$ for a given observed value $x$ can be given by $\overline{x}^{*}=x+\beta+\sqrt{2\beta x+\beta^2}$ and $\underline{x}^{*}=x-\frac{\beta}{2}-\sqrt{2\beta x+\frac{\beta^2}{4}}$ with a failure probability $\epsilon$, respectively. Note that we must have the lower bound $\underline{x}^{*}\geq0$. Utilizing a simple function transformation, the numerical result of upper bound $\overline{x}^{*}$ with Eq.~\eqref{eq20} is the same as (Eq.~\eqref{B7} in this paper) in Ref.~\cite{zhang2017improved}, while the analytic result of upper bound is more optimal in this work.
The numerical result of lower bound $\underline{x}^{*}$ with Eq.~\eqref{eq22} is different from in Ref.~\cite{zhang2017improved}, and the difference between two analytic results of lower bound is only $\beta$. However, we should point out that our result is always inferior to the Gaussian analysis, while the result of Ref.~\cite{zhang2017improved} is superior to the Gaussian analysis given a small observed value, details can be found in Fig.~\ref{f2}. It means that our result is rigorous while that of Ref.~\cite{zhang2017improved} is not. The case of small observed value is very important since the vacuum state is widely used in decoy-state method, especially for the experiment of measurement-device-independent QKD~\cite{Yin:2016:measurement}.

\section{Finite-key analysis for decoy-state BB84 QKD}
In this section, we exploit our statistical fluctuation analysis methods to deal with finite-key effects with composable security for the case of BB84 QKD with two decoy states.
Compared with previous results~\cite{curty2014finite,lim2014concise,zhang2017improved}, we provide the complete extractable secret key formula. For example,
the number of vacuum component events, the number of single-photon component events, and the phase error rate associated with the single-photons component events are all required to use observed values in the extractable secret key formula, while all or part of them are taken as the expected values in Ref.~\cite{curty2014finite,lim2014concise,zhang2017improved}. Obviously, they are observed values, for instance, the QKD system with single-photon source~\cite{tomamichel2012tight}.

\subsection{Protocol description}
The asymmetric coding BB84 protocol, based on which we consider our protocol, means that the bases $\mathsf{Z}$ and $\mathsf{X}$ are chosen with biased probabilities, both when Alice prepare the quantum states and when Bob measure those states.  Furthermore, intended to simplifying the protocol a little, we let the secret key be extracted only if Alice and Bob both choose the $\mathsf{Z}$ basis. Also, for the same purpose, the protocol will be built on the transmission of phase-randomized laser pulses and makes use of vacuum and weak decoy states. Below we provide a detailed description of the protocol with active basis choosing.
\newline

\noindent{\it{1.~Preparation.}}~The first three steps are repeated by Alice and Bob for $i=1,\ldots,N$ until the conditions
in the reconciliation step are satisfied. Alice will prepare weak coherent pulse and encode under the $\{\mathsf{Z},\mathsf{X}\}$ basis, along with an intensity $k \in \{\mu,\nu,0 \}$. Let the probability of choosing $\mathsf{Z}$ and $\mathsf{X}$ basis be $p_z$ and $p_x=1-p_z$. Simultaneously, the probabilities of selecting intensities are $p_{\mu}$, $p_{\nu}$ and $p_{0}=1-p_{\mu}-p_{\nu}$, respectively. Then Alice sends the weak coherent pulse to Bob through the insecure quantum channel.
\newline

\noindent{\it{2.~Measurement.}}~When receiving the pulse, Bob also chooses a basis $\mathsf{Z}$ and $\mathsf{X}$ with probabilities $q_{\rm z}$ and $q_{\rm x}=1-q_{\rm z}$, respectively. Then, he measures the state with two single-photon detectors in that basis. An effective event represents at least one detector click.
For double detector click event, he acquires a random bit value.
\newline

\noindent{\it{3.~Reconciliation.}}~Alice and Bob share the effective event, basis and intensity information with each other using an authenticated classical channel. We use the following sets $\cZ_{k}$ ($\cX_{k}$), which identifies signals where both Alice and Bob select the basis $\mathsf{Z}$ ($\mathsf{X}$) for $k$ intensity.
Then, they check for $|\cZ_{k}| \geq n^{\sZ}_{k}$ and  $|\cX_{k}| \geq n^{\sX}_{k}$  for all values of $k$.
They repeat step 1 to step 3 until these conditions are satisfied. We remark that the vacuum state prepared by Alice has no basis information.
\newline

\noindent{\it{4.~Parameter estimation.}}~After reconciling the basis and intensity choices, Alice and Bob will select a size of $n^{\sZ}=n^{\sZ}_{\mu}+n^{\sZ}_{\nu}$ to get a raw key pair $(\rv{Z}_{\rm A},\rv{Z}_{\rm B})$. All sets are used to compute the number of vacuum events $s_0^\sZ$ and single-photon events $s_1^\sZ$ and the phase error rate of single-photon events $\phi_1^\sZ$ in $\rv{Z}_{\rm A}$. After that, a condition should be met that the phase error rate $\phi_1^\sZ$ is less than $ \phi_{\rm{tol}}$, where $\phi_{\rm{tol}}$ is a predetermined phase error rate. If not, Alice and Bob abort the results and get started again. Otherwise, they move on to step 5.
\newline

\noindent{\it{5.~Postprocessing.}}
~First, Alice and Bob operate an error correction, where they reveal  at most $\lambda_{\rm EC}$ bits of information. Then, an error-verification step is performed using a random universal$_{2}$ hash function that announces $\lceil\log_2\frac{1}{\varepsilon_{\rm cor}}\rceil$ bits of information~\cite{wegman1981new}, where $\varepsilon_{\rm cor}$ is the probability that a pair of nonidentical keys passes the error-verification step. At last, there is a privacy amplification on their keys to get a secret key pair ($\rvS_{\rm A}$,$\rvS_{\rm B}$), both of which are $\ell$ bits, by using a random universal$_{2}$ hash function.
\newline

\subsection{Security bounds}
~Before stating how to calculate the security bound, we will spell out our security criteria, i.e., the so-called universally composable framework~\cite{muller2009composability}.
We have two criteria ($\varepsilon_{\rm cor}$ and $\varepsilon_{\rm sec}$) to determine how secure of our protocol. If $\Pr[\rv{S}_{\rm A}\not=\rv{S}_{\rm B}] \leq \varepsilon_{\rm cor}$, which means the secret keys are identical except with a small probability $\varepsilon_{\rm cor}$, we can call it is $\varepsilon_{\rm cor}$-correct. Meanwhile, if $(1-p_{\rm abort})\|\rho_{\rm AE}-U_{\rm A} \otimes \rho_{\rm E}\|_1/2 \leq \varepsilon_{\rm sec}$, we can call it is $\varepsilon_{\rm sec}$-secret. Thereinto, $\rho_{\rm AE}$ is the classical-quantum state describing the joint state of $\rv{S}_{\rm A}$ and $\rv{E}$,~$U_{\rm A}$ is the uniform mixture of all possible values of $\rv{S}_{\rm A}$,~and $p_{\rm abort}$ is the probability that the protocol aborts. This security criterion guarantees that the pair of secret keys can be unconditionally safe to use, we can call the protocol is $\varepsilon$-secure if it is $\varepsilon_{\rm cor}$-correct and $\varepsilon_{\rm sec}$-secret with $\varepsilon_{\rm cor}+\varepsilon_{\rm sec}\leq \varepsilon$.

The protocol is $\varepsilon_{\rm sec}$-secret if the secret key of length $\ell$ satisfies
\begin{equation}\label{eq27}
\begin{split}
\ell = &\underline{s}_0^\sZ+\underline{s}_1^\sZ\left[1-h\left( \overline{\phi}_1^\sZ \right)\right]-\lambda_{\rm EC}-\log_2\frac{2}{\varepsilon_{\rm cor}}-6\log_2\frac{23}{\varepsilon_{\rm sec}},
\end{split}
\end{equation}
where~$h(x):=-x\log_2x-(1-x)\log_2(1-x)$~is the binary Shannon entropy function. Note that observed values $\underline{s}_0^\sZ$, $\underline{s}_1^\sZ$ and $\overline{\phi}_1^\sZ$ are the lower bound for the number of vacuum events, the lower bound for the number of single-photon events, and the upper bound for the phase error rate associated with the single-photons events in $\rv{Z}_{\rm A}$, respectively. Here, we simply assume an error correction leakage $\lambda_{\rm EC} = n^{\sZ}\zeta h(E^{\sZ})$, with the efficiency of error correction $\zeta = 1.22$ and the bit error rate $E^{\sZ}$ in $(\rv{Z}_{\rm A},\rv{Z}_{\rm B})$.

Let $n^{\sZ}_{k}$ and  $n^{\sX}_{k}$  are the observed number of bit in set $\cZ_{k}$ and $\cX_{k}$. Let $m^{\sZ}_{k}$ and $m^{\sX}_{k}$ denote the observed number of bit error in set $\cZ_{k}$ and $\cX_{k}$. Note that one cannot obtain the $m^{\sZ}_{\mu}$ and $m^{\sZ}_{\nu}$, which we just hypothetically use to estimate the error correction information. The bit error rate is $E^{\sZ}=(m^{\sZ}_{\mu}+m^{\sZ}_{\nu})/n^{\sZ}$. By using the decoy-state method for finite sample sizes, we can have the lower bound on the expected numbers of vacuum event $\underline{s}_0^{\sZ^*}$ and single-photon event $\underline{s}_1^{\sZ^{*}}$ in $\rv{Z}_{\rm A}$,
\begin{equation}
\begin{aligned}\label{eq1}
\underline{s}_0^{\sZ^*}\geq &( e^{-\mu}p_\mu+ e^{-\nu}p_\nu) \frac{\underline{n}_0^{\sZ^*}}{p_0},\\
\underline{s}_1^{\sZ^*}\geq&\frac{\mu^{2} e^{-\mu}p_\mu+\mu\nu e^{-\nu}p_\nu}{\mu\nu-\nu^2}\\
&\times\left(e^\nu \frac{\underline{n}_\nu ^{\sZ^*}}{p_\nu}-\frac{\nu^2}{\mu^2}e^\mu \frac{\overline{n}_\mu^{\sZ^*}}{p_\mu}-\frac{\mu^2-\nu^2}{\mu^2}\frac{\overline{n}_0^{\sZ^*}}{p_0}\right),
\end{aligned}
\end{equation}
where $\underline{n}_0^{\sZ^*}$ and $\underline{n}_\nu ^{\sZ^*}$ ($\overline{n}_\mu^{\sZ^*}$ and $\overline{n}_0^{\sZ^*}$) are the lower (upper) bound of expected values associated with the observed values $n_0^{\sZ}$ and $n_\nu ^{\sZ}$ ($n_\mu^{\sZ}$ and $n_0^{\sZ}$). We can also calculate the lower bound on the expected number of single-photon event $\underline{s}_1^{\sX^{*}}$ and the upper bound on the expected number of bit error $\underline{t}_1^{\sX^{*}}$ associated with the single-photon event in $\cX_{\mu}\cup\cX_{\nu}$,
\begin{equation}
\begin{aligned}\label{eq1}
\underline{s}_1^{\sX^*}\geq&\frac{\mu^{2} e^{-\mu}p_\mu+\mu\nu e^{-\nu}p_\nu}{\mu\nu-\nu^2}\\
&\times\left(e^\nu \frac{\underline{n}_\nu ^{\sX^*}}{p_\nu}-\frac{\nu^2}{\mu^2}e^\mu \frac{\overline{n}_\mu^{\sX^*}}{p_\mu}-\frac{\mu^2-\nu^2}{\mu^2}\frac{\overline{n}_0^{\sX^*}}{p_0}\right),\\
\overline{t}_1^{\sX^{*}}\leq& \frac{\mu e^{-\mu}p_{\mu}+\nu e^{-\nu}p_{\nu}}{\nu}\left(e^{\nu}\frac{\overline{m}_{\nu}^{\sX^{*}}}{p_{\nu}}-\frac{\underline{n}_{0}^{\sX^{*}}}{2p_{0}}\right),
\end{aligned}
\end{equation}
where we use a fact that expected value $m_{0}^{\sX^*}\equiv n_{0}^{\sX^*}/2$. Parameters $\underline{n}_0^{\sX^*}$ and $\underline{n}_\nu ^{\sX^*}$ ($\overline{n}_\mu^{\sX^*}$, $\overline{n}_0^{\sX^*}$ and $\overline{m}_{\nu}^{\sX^{*}}$) are the lower (upper) bound of expected values associated with the observed values $n_0^{\sX}$ and $n_\nu ^{\sX}$ ($n_\mu^{\sX}$, $n_0^{\sX}$ and $m_{\nu}^{\sX}$). The nine expected values $\underline{n}_0^{\sZ^*}$, $\underline{n}_\nu ^{\sZ^*}$, $\overline{n}_\mu^{\sZ^*}$, $\overline{n}_0^{\sZ^*}$, $\underline{n}_0^{\sX^*}$, $\underline{n}_\nu ^{\sX^*}$, $\overline{n}_\mu^{\sX^*}$, $\overline{n}_0^{\sX^*}$ and $\overline{m}_{\nu}^{\sX^{*}}$ can be obtained by using the variant of Chernoff bound with Eq.~\eqref{eq25} and Eq.~\eqref{eq26} for each parameter with failure probability $\varepsilon_{\rm sec}/23$, for example, $\underline{n}_\nu ^{\sZ^*}=n_\nu ^{\sZ}-\Delta^{L}(n_\nu ^{\sZ},\varepsilon_{\rm sec}/23)$.

Once acquiring the four expected values $\underline{s}_0^{\sZ^*}$, $\underline{s}_1^{\sZ^*}$, $\underline{s}_1^{\sX^*}$ and $\overline{t}_1^{\sX^{*}}$, one can exploit the Chernoff bound with Eq.~\eqref{eq17} and Eq.~\eqref{eq18} to calculate the corresponding observed values $\underline{s}_0^{\sZ}$, $\underline{s}_1^{\sZ}$, $\underline{s}_1^{\sX}$ and $\overline{t}_1^{\sX}$ for each parameter with failure probability $\varepsilon_{\rm sec}/23$, for example, $\underline{s}_1^{\sZ}=\underline{s}_1^{\sZ^*}(1-\delta^{L}(\underline{s}_1^{\sZ^*},\varepsilon_{\rm sec}/23))$.
By using the random sampling without replacement with Eq.~\eqref{eq5}, one can calculate the upper bound of hypothetically observed phase error rate associated with the single-photon events in $\rv{Z}_{\rm A}$,
\begin{equation}\label{eq28}
\overline{\phi}_1^\sZ=\frac{\overline{t}_1^\sX}{\underline{s}_1^\sX}+\gamma^{U}\left(\underline{s}_1^\sZ,\underline{s}_1^\sX,\frac{\overline{t}_1^\sX}{\underline{s}_1^\sX},\frac{\varepsilon_{\rm sec}}{23}\right).
\end{equation}

\subsection{Simulation}

In order to show the performance of our method in terms of the secret key rate and the secure transmission distance, we consider a fiber-based QKD system model with active basis choosing measurement. We use the widely used parameters of a practical QKD system~\cite{gobby2004quantum}, as listed in Table I. For a given experiment, one can directly acquire the parameters $n_{k}^{\sZ}$, $n_{k}^{\sX}$, $m_{k}^{\sZ}$ and $m_{k}^{\sX}$.
For simulation, we can use the formulas $n_k^\sZ=Np_{k}p_{z}q_{z} Q_k^\sZ$, $n_k^\sX=Np_{k}p_{x}q_{x} Q_k^\sX$,
$m_k^\sZ=Np_{k}p_{z}q_{z} E_k^\sZ Q_k^\sZ$ and $m_k^\sX=Np_{k}p_{x}q_{x} E_k^\sX Q_k^\sX$,
where $Q_{k}^{\sZ}$ and $Q_{k}^{\sX}$ are the gain of $Z$ and $X$ basis when Alice chooses optical pulses with intensity $k$.
For vacuum state without basis information, we can reset $n_0^\sZ=Np_{0}q_{z} Q_0^\sZ$, $n_0^\sX=Np_{0}q_{x} Q_0^\sX$,
$m_0^\sZ=Np_{0}q_{z} E_0^\sZ Q_0^\sZ$ and $m_0^\sX=Np_{0}q_{x} E_0^\sX Q_0^\sX$.
$E_{k}^{\sZ}$ and $E_{k}^{\sX}$ are the bit error rate of $Z$ and $X$ basis when Alice chooses optical pulses with intensity $k$.  Without loss of generality, these gain and bit error rate parameters can be given by~\cite{ma2005practical}
\begin{equation}
\begin{aligned}\label{}
&Q_{k}^\sZ=Q_{k}^\sX=1-(1-Y_{0})e^{-k\eta},\\
&E_{k}^{\sZ}Q_{k}^\sZ=E_{k}^{\sX}Q_{k}^{\sX}=e_{d}Q_{k}^\sZ+(e_0-e_d)Y_{0},
\end{aligned}
\end{equation}
where we assume that those observed values for different parameters can be denotes by their asymptotic values without Eve's disturbance. $\eta=\eta_{d}\times10^{-\alpha L/10}$ is the overall efficiency with the fiber length $L$ and single-photon detector.

\begin{table}[ht]
\centering
\caption{List of simulation parameters. $\eta_{d}$ is the detection efficiency of single-photon detector, $\zeta$ is the efficiency of error correction, $\alpha$ is the attenuation coefficient of single-mode fiber, $e_{d}$ is the misalignment rate, and $N$ is the number of optical pulses sent by Alice.}
\begin{tabular}{cccccccc}
\hline
\hline
$\eta_{d}$ & $Y_{0}$ & $e_{d}$ & $\alpha$ & $\zeta$ & $\varepsilon_{\rm sec}$ & $\varepsilon_{\rm cor}$ & $N$\\
\hline
$4.5\%$ & $1.7\times10^{-6}$ & $3.3\%$ & $0.21~\textrm{dB/km}$ &$1.22$ &$10^{-10}$ & $10^{-15}$ &$10^{10}$
\\
\hline
\hline
\end{tabular}
\end{table}

\begin{figure}[t!]
  \centering
  \includegraphics[width=86mm]{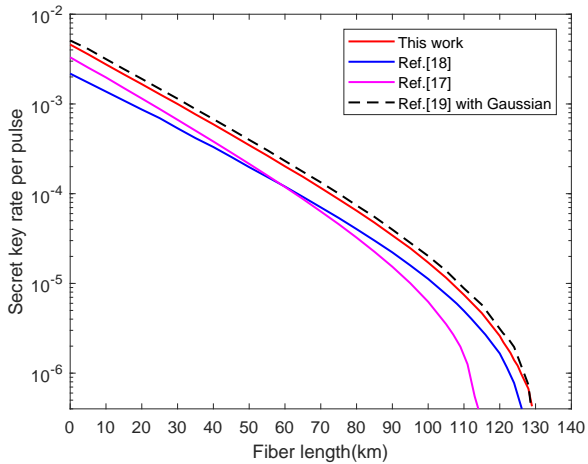}
  \caption{\textbf{The secret key rate vs fiber length. }\label{fig3} It shows the comparison of the secret key rates of different statistical fluctuation methods. Numerically optimized secret key rates with logarithmic scale are obtained for a predetermined signals $N=10^{10}$.}
\end{figure}

To show the advantage of our results compared with previous works~\cite{curty2014finite,lim2014concise,zhang2017improved}, we drew the curves about the secret key rate $\ell/N$ as function of the fiber length, as shown in Fig.~\ref{fig3}. For a given number of signals $10^{10}$, only ten seconds in 1 GHz system, we optimize numerically $\ell/N$ over all the free parameters. For fair comparison, we add a step about from expected value to observed value estimation for all curves, which is not taken into account in Refs.~\cite{curty2014finite,lim2014concise}.
The corresponding methods of Refs.~\cite{ma2005practical,curty2014finite,lim2014concise,zhang2017improved} to deal with statistical fluctuation can be summarized in Appendix B. Note that the black dashed line uses the Gaussian analysis to obtain expected value instead of the inverse solution Chernoff bound method~\cite{zhang2017improved}.
The simulation results show that the secret key rate and secure transmission distance of our method have significant advantage under the security against the general attacks.

\section{Conclusion}
In this work, we proposed the almost optimal analytical formulas to deal with the statistical fluctuation under the security against the general attacks. Analytical formulas of classical postprocessing can be expediently used in practical system, which do not introduce complex calculations of resource consumption. Our methods can directly increase the performance without changing the quantum process, which should be widely used to quantum cryptography protocols against the finite-size effects. In order to compare with previous works, we establish the complete finite-key analysis for decoy-state BB84 QKD, including from observed value to expected value, from expected value to observed value and from the observed bit error of $\mathsf{X}$ basis to hypothetical observed phase error of $\mathsf{Z}$ basis. We remark that the joint constraint method~\cite{zhou2016making} can further decrease the statistical fluctuation. However, we do not consider this issue in this paper due to without the analytical solutions, which is difficult to implement in commercial systems.

\section{Acknowledgements}

We gratefully acknowledges support from the National Natural Science Foundation of China under Grant No. 61801420, the Fundamental Research Funds for the Central Universities and the Nanjing University.

\appendix

\section{Proof of random sampling without replacement}

\emph{Proof.} Here, we use the technique of Ref.~\cite{korzh2015provably} to acquire the correct analytical results. We remark that the result of Ref.~\cite{korzh2015provably} is wrong due to the incorrect inequality scaling about binomial coefficient and Eq. (11) in supplementary information of Ref.~\cite{korzh2015provably}.
For the upper tail, the failure probability $\epsilon$ can be bound by~\cite{fung2010practical,korzh2015provably,yin2019finite} $C_{k}^{k \lambda} C_{n}^{n \chi}/C_{n+k}^{(n+k)y}$, where $y=\lambda+\frac{n}{n+k} \gamma$ and $\chi=\lambda+\gamma$.
Let $F(\alpha, n)=\frac{\alpha^{-\alpha n}(1-\alpha)^{-(1-\alpha) n}}{\sqrt{2 \pi n \alpha(1-\alpha)}}$ and  using the sharp double inequality for binomial coefficient~\cite{stanica2001good}
\begin{equation}\label{eq52}
e^{-\frac{1}{8 \alpha n}} F(\alpha, n)<C_{n}^{\alpha n}<e^{\left(\frac{1}{12 n}-\frac{1}{12 n \alpha+1}-\frac{1}{12 n (1-\alpha)+1}\right)} F(\alpha, n),
\end{equation}
we can give the following inequality for failure probability
\begin{equation}\label{eq53}
\begin{split}
\frac{C_{k}^{k \lambda} C_{n}^{n \chi}}{C_{n+k}^{(n+k)y}}<&\frac{e^{\ln2 \cdot\left[n h\left(\chi\right)+k h\left(\lambda\right)-(n+k) h\left(y\right)\right]}}{\sqrt{2 \pi n k \lambda(1-\lambda) /(n+k)}}\sqrt{\frac{y(1-y) }{\chi(1-\chi)}}\\
&\times e^{\left(\frac{1}{8(n+k)y}+\frac{1}{12k}-\frac{1}{12k\lambda+1}-\frac{1}{12k(1-\lambda)+1}\right)}\\
&\times e^{\left(\frac{1}{12 n}-\frac{1}{12 n\chi+1}-\frac{1}{12 n(1-\chi)+1}\right)},
\end{split}
\end{equation}
where Shannon entropy function $h(x)=-x \log _{2} x-(1-x) \log _{2}(1-x)$.
Note that one can prove $e^{\left(\frac{1}{8(n+k)y}+\frac{1}{12k}-\frac{1}{12k\lambda+1}-\frac{1}{12k(1-\lambda)+1}+\frac{1}{12 n}-\frac{1}{12 n\chi+1}-\frac{1}{12 n(1-\chi)+1}\right)}<1$ and $\sqrt{\frac{y(1-y) }{\chi(1-\chi)}}<1$ for $n,k>0$ and $0<\lambda<y<\chi\leq0.5$.
Thereby, the inequality can be given by
\begin{equation}\label{eq53}
\begin{split}
\frac{C_{k}^{k \lambda} C_{n}^{n \chi}}{C_{n+k}^{(n+k)y}}<&\frac{e^{\ln2 \cdot\left[n h\left(\chi\right)+k h\left(\lambda\right)-(n+k) h\left(y\right)\right]}}{\sqrt{2 \pi n k \lambda(1-\lambda) /(n+k)}}.
\end{split}
\end{equation}
By using Taylor expanding for the case of $n\geq k$, we have
$n h(\chi)+k h(\lambda)-(n+k) h(y) \leq \frac{h^{\prime \prime}(y)}{2} \frac{\gamma^{2} n k}{n+k}$,
where $h^{\prime \prime}(y)=-\frac{1}{y(1-y) \ln 2}$.
Therefore, by solving a quadratic equation with one unknown, we have
\begin{equation}\label{eq59}
\gamma=\frac{\frac{(1-2 \lambda) n G}{n+k}+\sqrt{\frac{n^{2} G^{2}}{(n+k)^{2}}+4 \lambda (1-\lambda)G}}{2+2 \frac{n^{2} G}{(n+k)^{2}}},
\end{equation}
where parameter $G=\frac{n+k}{nk}\ln{\frac{n+k}{2\pi nk\lambda(1-\lambda)\epsilon^{2}}}$.
By using Taylor expanding for the case of $n\leq k$, we have the following inequalities
$n h(\chi)+k h(\lambda)-(n+k) h(y)\leq n h(\lambda)+k h(\chi)-(n+k) h(z)\leq \frac{h^{\prime \prime}(z)}{2} \frac{\gamma^{2} n k}{n+k}$
where $z=\lambda+\frac{k}{n+k} \gamma$ and $h^{\prime \prime}(z)=-\frac{1}{z(1-z) \ln 2}$. Therefore, by solving a quadratic equation with one unknown, we have
\begin{equation}\label{eq63}
\gamma=\frac{\frac{(1-2 \lambda) k G}{n+k}+\sqrt{\frac{k^{2} G^{2}}{(n+k)^{2}}+4 \lambda(1-\lambda)G}}{2+2 \frac{k^{2} G}{(n+k)^{2}}}.
\end{equation}
Note that the above result is always true for all $n,k>0$ and $0<\lambda<\chi\leq0.5$.

\section{Previous methods}
In this section, we summarize the statistical fluctuation method used previously as follows.

\subsection{Method in Ref.~\cite{curty2014finite}}
The upper bound of the random sampling without replacement can be calculated by using the Serfling inequality,
\begin{equation}
\begin{aligned}
\gamma^{U}=\sqrt{\frac{(n+k)(k+1)}{nk^{2}}\ln\epsilon^{-1}}.
\end{aligned}
\end{equation}

The upper bound and lower bound of expected value for a given observed value can be calculated by using the multiplicative form Chernoff bound as follows.
We always can obtain the worst lower bound of expected value, $\mu_{\rm L}=x-\sqrt{N/2\ln{\epsilon^{-1}}}$, where $N$ is the total number of random variables.
Let $test_1$, $test_2$ and $test_3$ denote, respectively,
the following three conditions: $\mu_{\rm L}\geq\frac{32}{9}\ln(2\epsilon_{1}^{-1})$,
$\mu_{\rm L}>3\ln\epsilon_{2}^{-1}$ and $\mu_{\rm L}>\left(\frac{2}{2e-1}\right)^{2}\ln\epsilon_{2}^{-1}$,
and let $g(x,y)=\sqrt{2x\ln{y^{-1}}}$. Now:
\begin{enumerate}
\item When $test_1$ and $test_2$ are fulfilled, we have that
$\Delta^{U}=g(x, \epsilon_{1}^4/16)$ and
$\Delta^{L}=g(x, \epsilon_{2}^{3/2})$.
\item When $test_1$ and $test_3$ are fulfilled (and $test_2$ is not fulfilled), we have that
$\Delta^{U}=g(x, \epsilon_{1}^4/16)$ and
$\Delta^{L}=g(x, \epsilon_{2}^{2})$.
\item When $test_1$ is fulfilled and $test_3$ is not fulfilled, we have that
$\Delta^{U}=g(x, \epsilon_{1}^4/16)$ and $\Delta^{L}=\sqrt{N/2\ln{\epsilon_{2}^{-1}}}$.
\item When
$test_1$ is not fulfilled and $test_2$ is fulfilled, we have that
$\Delta^{U}=\sqrt{N/2\ln{\epsilon_{1}^{-1}}}$ and
$\Delta^{L}=g(x, \epsilon_{2}^{3/2})$.
\item When $test_1$ and $test_2$ are not fulfilled, and $test_3$ is fulfilled, we have that
$\Delta^{U}=\sqrt{N/2\ln{\epsilon_{1}^{-1}}}$ and
$\Delta^{L}=g(x, \epsilon_{2}^{2})$.
\item When $test_1$, $test_2$ and $test_3$ are not fulfilled, we have that
$\Delta^{U}=\sqrt{N/2\ln{\epsilon_{1}^{-1}}}$ and $\Delta^{L}=\sqrt{N/2\ln{\epsilon_{2}^{-1}}}$
\end{enumerate}
To simplify this simulation, we consider the case of $\epsilon=\epsilon_{1}=\epsilon_{2}$. For all observed value $x$, we make $\overline{x}^{*}=x+\Delta^{U}$ and $\underline{x}^{*}=x-\Delta^{L}$, where
\begin{equation}
\begin{aligned}\label{eqB2}
&\Delta^{U}=\sqrt{8\beta x+8x\ln2},\\
&\Delta^{L}=\sqrt{3\beta x}.\\
\end{aligned}
\end{equation}
Note that it is not rigorous in Eq.~\eqref{eqB2} for small $x$.

\subsection{Method in Ref.~\cite{lim2014concise}}
The upper bound of the random sampling without replacement can be calculated by
\begin{equation}
\gamma^{U}=\sqrt{\frac{(n+k)\lambda(1-\lambda)}{nk\ln2}\log_{2}{\frac{n+k}{nk\lambda(1-\lambda)\epsilon^{2}}}},
\end{equation}
where the result is true only when $n$ and $k$ are large.

The upper bound and lower bound of expected value for a given observed value can be calculated by using the tailored Hoeffding inequality for decoy-state method. Let $x_{k}$ be the observed value for $k$ intensity and $X=\sum_{k}x_{k}$. Therefore, we have $\overline{x}_{k}^{*}=x_{k}+\Delta^{U}$ and $\underline{x}_{k}^{*}=x_{k}-\Delta^{L}$, where
\begin{equation}
\begin{aligned}\label{}
\Delta^{U}=\Delta^{L}=\sqrt{X/2\ln\epsilon^{-1}}.
\end{aligned}
\end{equation}
Note that the deviation is the same for all intensities of $k$, which will lead large fluctuation for small intensity, especially vacuum state.

\subsection{Method in Ref.~\cite{zhang2017improved}}
The upper bound of the random sampling without replacement can be calculated by using the following transcendental equation,
\begin{equation}
\begin{aligned}\label{}
h&\left(\lambda+\frac{n}{n+k}\gamma^{U}\right)-\frac{k}{n+k}h(\lambda)-\frac{n}{n+k}h(\lambda+\gamma^{U})\\
&=\frac{1}{2(n+k)}\log_{2}\frac{n+k}{nk\lambda(1-\lambda)\epsilon^{2}}.
\end{aligned}
\end{equation}

The upper bound and lower bound of expected value for a given observed value can be calculated by using the Gaussian analysis. Therefore, we have $\overline{x}^{*}=x+\Delta^{U}$ and $\underline{x}^{*}=x-\Delta^{L}$ with
\begin{equation}
\begin{aligned}\label{}
\Delta^{U}=\Delta^{L}={\rm erfcinv}(2\epsilon)\sqrt{2x} ,
\end{aligned}
\end{equation}
where $a={\rm erfcinv}(b)$ is the inverse function of $b={\rm erfc}(a)$ and ${\rm erfc}(a)=\frac{2}{\sqrt{\pi}}\int_{a}^{\infty}e^{-t^2}dt$ is the complementary error function.

Furthermore, the upper bound and lower bound of expected value for a given observed value can also be calculated by using the inverse solution Chernoff bound. Therefore, we have $\overline{x}^{*}=x/(1-\delta^{U})$ and $\underline{x}^{*}=x/(1+\delta^{L})$, where $\delta^{U}$ and $\delta^{L}$ can be obtained by using the following transcendental equation,
\begin{equation}
\begin{aligned}\label{B7}
\frac{x}{1-\delta^{U}}[-\delta^{U}-(1-\delta^{U})\ln(1-\delta^{U})]=\ln\epsilon,\\
\frac{x}{1+\delta^{L}}[\delta^{L}-(1+\delta^{L})\ln(1+\delta^{L})]=\ln\epsilon,\\
\end{aligned}
\end{equation}
while the slightly looser analytic result can be given by
\begin{equation}
\begin{aligned}\label{}
\delta^{U}=\frac{\sqrt{8\beta x+9\beta^{2}}-\beta}{2(x+\beta)},\\
\end{aligned}
\end{equation}
and
\begin{equation}
\begin{aligned}\label{}
\delta^{L}=\frac{\sqrt{8\beta x+\beta^{2}}+3\beta}{2(x-\beta)}.\\
\end{aligned}
\end{equation}
Through simple calculation, the upper bound and lower bound are $\overline{x}^{*}=x+\frac{3}{2}\beta+\sqrt{2\beta x+\frac{9}{4}\beta^{2}}$ and $\underline{x}^{*}=x+\frac{\beta}{2}-\sqrt{2\beta x+\frac{\beta^{2}}{4}}$, respectively.

\bibliographystyle{apsrev}



\end{document}